\documentclass{nature}

\usepackage{graphicx}
\usepackage{dcolumn}
\usepackage{bm}
\usepackage{xspace}
\usepackage{hyperref}
\usepackage{ifthen}
\usepackage[nomessages]{fp}
\usepackage{amssymb}
\usepackage{amsmath}
\usepackage{color}
\usepackage[normalem]{ulem}
\usepackage{physics}

\newcommand{\bkbo}{\ifmmode\text{Ba}_{1-x}\text{K}_{x}\text{BiO}_{3}\else$\text{Ba}_{1-x}\text{K}_{x}\text{BiO}_{3}$\xspace\fi}
\newcommand{\bpbo}{\ifmmode\text{BaPb}_{1-x}\text{Bi}_{x}\text{O}_{3}\else$\text{BaPb}_{1-x}\text{Bi}_{x}\text{O}_{3}$\xspace\fi}
\newcommand{\bbo}{\ifmmode\text{BaBiO}_{3}\else$\text{BaBiO}_{3}$\xspace\fi}
\newcommand{\tc}{\ifmmode{T}_{c}\else${T}_{c}$\xspace\fi}
\newcommand{\tp}{\ifmmode{T}_{p}\else${T}_{p}$\xspace\fi}
\newcommand{\ef}{\ifmmode{E}_{F}\else${E}_{F}$\xspace\fi}
\newcommand{\kf}{\ifmmode{k}_{F}\else${k}_{F}$\xspace\fi}

\makeatletter
\let\saved@includegraphics\includegraphics
\AtBeginDocument{\let\includegraphics\saved@includegraphics}
\renewenvironment*{figure}{\@float{figure}}{\end@float}
\makeatother

\title{Cooling a polaronic liquid: Phase mixture and pseudogap-like spectra in superconducting $\bm{\bkbo}$}

\author{M.~Naamneh$^{1}$, M.~Yao$^{1}$, J.~Jandke$^{1}$, J.~Ma$^{1}$, Z.~Risti\'{c}$^{1}$, J.~Teyssier$^{2}$, A.~Stucky$^{2}$, D.~van der Marel$^{2}$, D.~J.~Gawryluk$^{3}$\footnote{On leave from Institute of Physics, Polish Academy of Sciences, Aleja Lotnikow 32/46, PL-02-668 Warsaw, Poland}, T.~Shang$^{3}$, M.~Medarde$^{3}$, E.~Pomjakushina$^{3}$, S.~Li$^{4}$, T. Berlijn$^{5,6}$, S.~Johnston$^{4,7}$, M.~M\"uller$^{8}$, J.~Mesot$^{9}$, M.~Shi$^{1}$, M.~Radovi\'{c}$^{1}$ \& N.~C.~Plumb$^{1}$}

\begin{document}

\sloppy

\maketitle

\begin{affiliations}
  \item Photon Science Division, Paul Scherrer Institut, CH-5232 Villigen PSI, Switzerland
  \item Department of Quantum Matter Physics (DQMP), University of Geneva, 24 quai Ernest-Ansermet, 1211 Geneva 4, Switzerland
  \item Laboratory for Multiscale Materials Experiments, Paul Scherrer Institut, CH-5232 Villigen-PSI, Switzerland
  \item Department of Physics and Astronomy, University of Tennessee, Knoxville, Tennessee 37996-1200, USA
  \item Center for Nanophase Materials Sciences, Oak Ridge National Laboratory, Oak Ridge, Tennessee 37831, USA
  \item Computational Science and Engineering Division, Oak Ridge National Laboratory, Oak Ridge, Tennessee 37831, USA
  \item Joint Institute for Advanced Materials at The University of Tennessee, Knoxville, Tennessee 37996, USA 
  \item Condensed Matter Theory Group, Paul Scherrer Institut, CH-5232 Villigen PSI, Switzerland
  \item Paul Scherrer Institut, CH-5232 Villigen PSI, Switzerland
\end{affiliations}

\begin{abstract}
Many complex electronic systems exhibit so-called pseudogaps, which are poorly-understood suppression of low-energy spectral intensity in the absence of an obvious gap-inducing symmetry. Here we investigate the superconductor $\bm{\bkbo}$ near optimal doping, where unconventional transport behavior and evidence of pseudogap(s) have been observed above the superconducting transition temperature $\bm{\tc}$, and near an insulating phase with long-range lattice distortions. Angle-resolved photoemission spectroscopy (ARPES) reveals a dispersive band with vanishing quasiparticle weight and ``tails'' of deep-energy intensity that strongly decay approaching the Fermi level. Upon cooling below a transition temperature $\bm{\tp > \tc}$, which correlates with a change in the slope of the resistivity vs. temperature, a partial transfer of spectral weight near $\bm{\ef}$ into the deep-binding energy tails is found to result from metal-insulator phase separation. Combined with simulations and Raman scattering, our results signal that insulating islands of ordered bipolarons precipitate out of a disordered polaronic liquid and provide evidence that this process is regulated by a crossover in the electronic mean free path.
\end{abstract} 

Pseudogaps represent a departure from the expectations of standard band theory and the Fermi liquid theory of electronic excitations, which together serve as a successful starting point for understanding many condensed matter systems. They could potentially originate from any ways in which the conventional theories might break down, e.g., due to disorder, fluctuations, strong interactions, and/or strong correlations. But it is also conceivable that some observed pseudogaps might be less mysterious than they first seem, in the sense that they are rooted a ``hidden'' order that, once revealed, could straightforwardly explain the opening of a gap. 

Pseudogaps are often observed in strongly correlated transition metal systems where the charge, spin, orbital, and lattice degrees of freedom can intertwine, leading to the emergence of exotic and poorly understood phases. These include ordered insulator or ``bad metal'' phases near half-filling (e.g., Mott, charge density wave, and spin density wave orders), metal-insulator transitions, superconductivity, anomalous transport behaviors (e.g., ``strange metal'' behavior and colossal magnetoresistance), intrinsic disorder and phase separation/fluctuation. Examples of such phenomena are found in diverse materials, including the cuprates\cite{Lee2006}, manganites\cite{Dagotto2005}, transition metal dichalcogenides\cite{Manzeli2017}, iron-based superconductors\cite{Si2016}, and rare earth nickelates\cite{Medarde1997}. 

\begin{figure}
    \centering
    \includegraphics[width=\columnwidth]{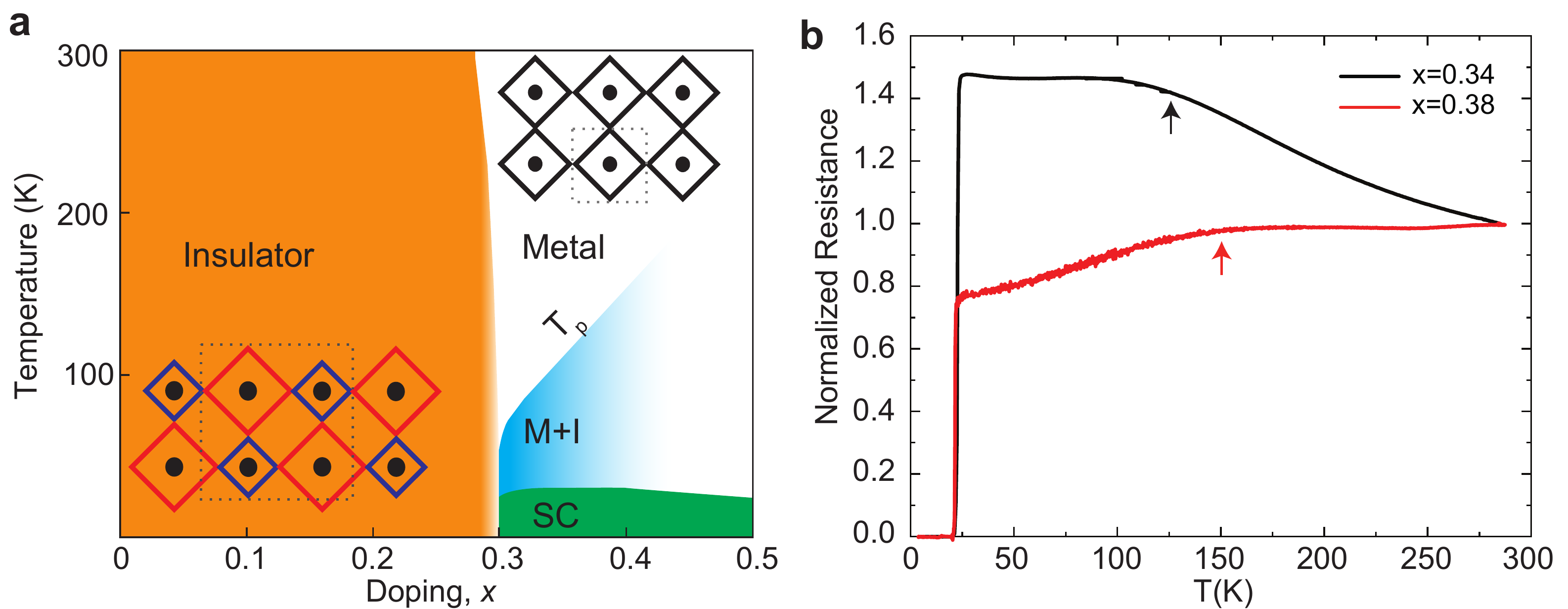}
    \caption{\textbf{Electronic properties of \bkbo.}
        \textbf{a}, Sketch of the doping- and temperature-dependent electronic phase diagram. The overlays are 2D schematics of the lattice illustrating ``breathing'' distortions essential to the insulating phase (left), as well as the idealized undistorted lattice in the metallic phase (right). Dots represent bismuth sites, while vertices are oxygen atoms. 
        \textbf{b}, The temperature dependence of the resistivity normalized by the room temperature resistance from films with $x=0.34$ and $x=0.38$. The arrows mark the temperature $T_p$ where the slope of the resistivity is changed. 
    }
    \label{fig:intro}
\end{figure}

In perovskite bismuth oxides such as \bkbo (BKBO), a complex phenomenology similar to that of transition metal oxides emerges out of a very different setting. Here, short-range Coulomb and spin interactions can largely be neglected\cite{Plumb2016}, but short-range electron-lattice interactions (polarons) play a crucial role, offering an opportunity to examine the signatures and origins of many-body phenomena from a new perspective. The phase diagram of BKBO is sketched in Fig.~\ref{fig:intro}\textbf{a}. The insulating phase at zero doping ($x=0$, half filling) is tied to a static long-range structural distortion. Here, alternating BiO$_6$ octahedra collapse in a three-dimensional breathing distortion, as illustrated schematically in the overlay of the phase diagram. Several studies have shown that the parent ground state is bond disproportionated, meaning that hole pairs are trapped in combinations of the O $2p$ orbitals in the collapsed octahedra, opening a gap in the predominantly oxygen-derived conduction band\cite{Shen1990,Ignatov1996,Foyevtsova2015,Plumb2016,Khazraie2018}. The localized charge pairs due to coupling to the lattice distortion may be regarded as a frozen lattice of bipolarons. This polaronic view of the parent compound's ground state has been widely adopted in theoretical approaches to understanding the perovskite bismuth oxides\cite{Rice1981,Micnas1990,Menushenkov2001,Franchini2009}.

BKBO transitions to a metal as a function of doping and/or temperature, and even becomes superconducting up to as high as 34 K in bulk\cite{Jones1989}. Electrical transport evolves rapidly and dramatically as a function of doping across the superconducting region of the phase diagram. Measurements on single crystals by Nagata \emph{et al}.\cite{Nagata1999} showed that at $x = 0.34$, resistivity $\rho$ rises with decreasing temperature down to about 75 K, where it saturates before \tc is reached. Although the high-$T$ behavior of the resistivity at this doping resembles an insulator, our ARPES results will show that the slope change or ``elbow'' in $\rho(T)$ above \tc should be viewed as a bad-to-better metal transition. At $x = 0.39$ --- close to what is generally cited as optimal doping\cite{Sleight2015} --- $\rho(T)$ is nearly linear above \tc, with a weak elbow centered roughly around 140 K. Higher doped samples exhibit a more conventional, upward-curving $\rho(T)$. 

The unusual transport above \tc in under-/optimally-doped superconducting samples appears to coincide with pseudogap-like spectral features. Optical conductivity studies of BKBO noted a suppression of the low-energy spectral weight at room temperature\cite{Karlow1993,Blanton1993}. Similar spectra were found in an analogous region of the phase diagram of the closely related superconducting compound \bpbo (BPBO)\cite{Tajima1985}. Separately, angle-integrated photoemission measurements performed on superconducting BKBO observed an energetically sharper, temperature-dependent pseudogap-like suppression of near-\ef spectral intensity\cite{Chainani2001}.

Some studies have interpreted the unusual transport and spectral properties of these materials as evidence of coexisting metallic and insulating phases\cite{Tajima1985,Nagata1999,Nicoletti2017}. Meanwhile a mixture of structural phases has been directly observed in superconducting BPBO compositions\cite{Climent-Pascual2011}. Based on transmission electron microscopy experiments, it appears that nanoscale regions of ordered distortions in BPBO tend to percolate into stripe-like formations\cite{Giraldo-Gallo2015}. This calls to mind predictions of 2D stacking ``slices'' of concentrated holes in BKBO\cite{Bischofs2002} and strongly suggests that phase mixing is intrinsic to these materials in certain doping ranges.
Overall, however, the ubiquity of structural phase mixing in superconducting bismuth oxides and its relationship to electronic structure and transport properties has remained unclear.

Here we use ARPES and Raman measurements to investigate the issues of possible pseudogaps and phase separation underlying BKBO's unusual transport characteristics. The ARPES experiments were performed \emph{in situ} on freshly-grown thin films of BKBO, allowing us to overcome longstanding sample and surface quality issues associated with single crystals that had prevented successful ARPES measurements up to now. Figure~\ref{fig:intro}\textbf{b} introduces typical resistance-vs.-temperature curves, $R(T)$, of our films grown by pulsed laser deposition (PLD) from ablation targets with $x=0.34$ and $x=0.38$. The curves are normalized to the resistance at 300 K. Aside from the lower superconducting transition temperatures in the films ($T_c = 22$ K) and slightly shifted temperatures of the elbow features relative to bulk BKBO, the $R(T)$ characteristics agree well with measurements of single crystals found in the literature\cite{Nagata1999}. We will later spectroscopically correlate the resistivity elbow with a transition temperature, $T_p$, to metal-insulator phase separation, thus defining the ``M+I'' region of the phase diagram in Fig.~\ref{fig:intro}\textbf{a}. In what follows, we focus on $x=0.34$ samples. 

\section*{Overview of ARPES spectra}

\begin{figure}
    \centering
    \includegraphics[width=0.6\textwidth]{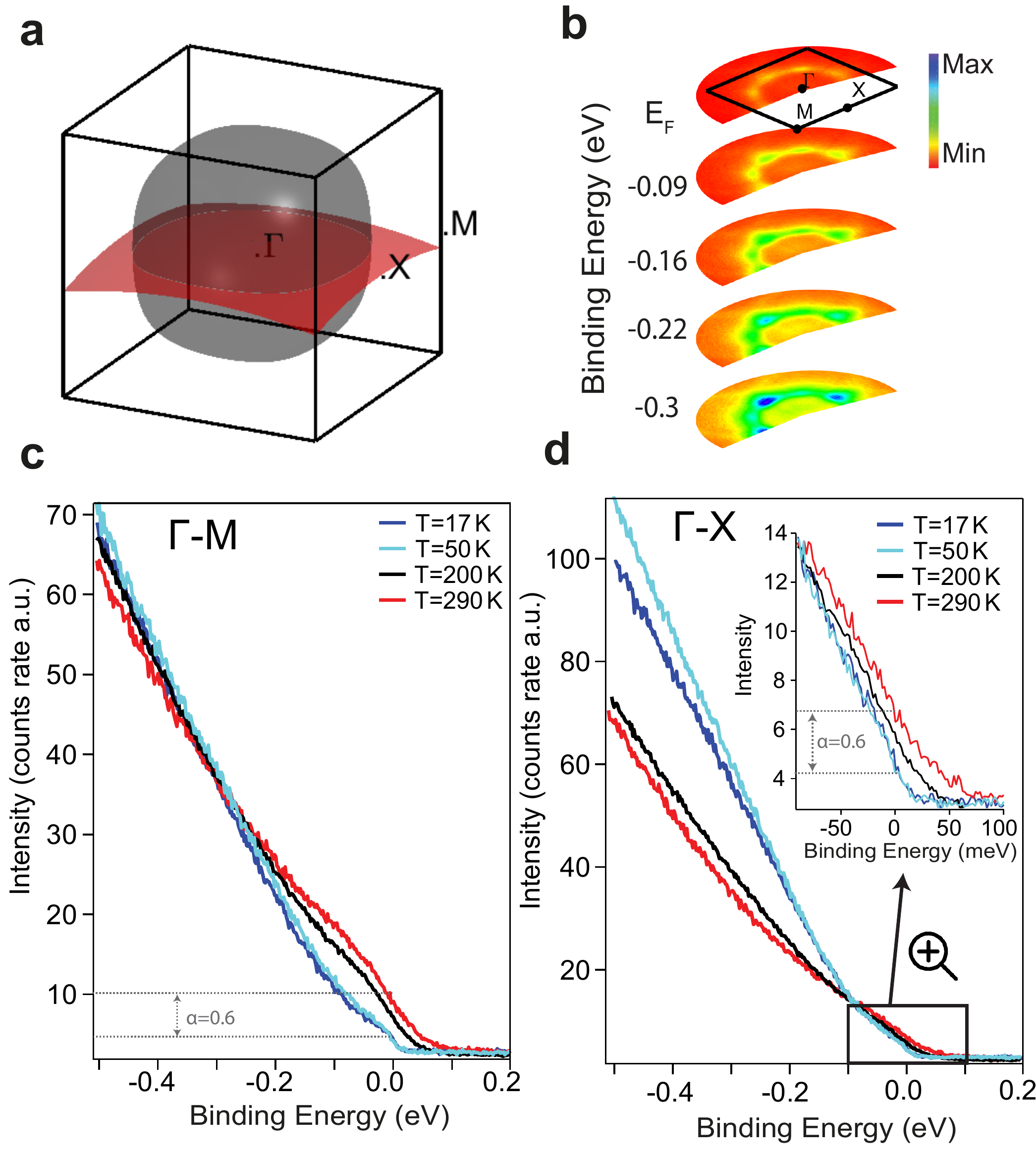}
    \caption{\textbf{Electronic structure of BKBO.}
        \textbf{a}, Brillouin zone of cubic BKBO drawn from a tight-binding model with filling corresponding to $x=0.34$ (0.66 e$^{-}$ per unit cell). ARPES data were acquired with an incident photon energy of 70 eV, corresponding to data points in momentum space indicated by the curved sheet, which is close to the $\Gamma-M-X$ plane. 
        \textbf{b}, ARPES intensity maps plotted as a function of binding energy. The sample temperature was 17 K. 
        \textbf{c}, \textbf{d} EDC's at different temperatures evaluated at \kf along the $\Gamma-M$ and $\Gamma-X$ directions, respectively. The inset of \textbf{d} highlights the energy region near \ef. The fraction of spectral intensity lost at \ef with decreasing temperature, denoted by $\alpha$, is equal in the two spectra. The EDC data is shown as-measured, without normalization. 
    }
    \label{fig:ARPES_EDC}
\end{figure}

Figure~\ref{fig:ARPES_EDC}\textbf{a} shows the Fermi surface of metallic, cubic BKBO with filling corresponding to $x=0.34$. The Fermi surface is drawn from a tight-binding model\cite{Sahrakorpi2000}, which is found to reasonably match the shape and volume of the Fermi surface obtained from ARPES measurements. ARPES constant energy maps at various binding energies are presented in Fig.~\ref{fig:ARPES_EDC}\textbf{b}. The measurements were performed using a photon energy of $h\nu=70$ eV, which corresponds to a sheet in momentum space that passes close to the $\Gamma-M-X$ plane, shown in Fig.~\ref{fig:ARPES_EDC}\textbf{a} (see Methods). The temperature was 17 K. 

The anomalous spectral lineshapes and first indications of pseudogap-like behavior in BKBO are evident from energy distribution curves (EDCs) at fixed momenta. EDCs at the Fermi momenta \kf along the $\Gamma-M$ and $\Gamma-X$ directions are shown in Figs.~\ref{fig:ARPES_EDC}\textbf{c} and \ref{fig:ARPES_EDC}\textbf{d}, respectively. The spectra taken at 290 K and 200 K exhibit Fermi steps at \ef but lack distinct quasiparticle peaks expected for a Fermi liquid. Instead, tails of spectral intensity rise monotonically to deeper binding energy. 

These peculiar EDC lineshapes give a spectroscopic view of the physics underlying the bad metallic transport behavior above \tc in BKBO near $x=0.34$. The electronic states lie far from the Fermi liquid regime in which sharp peaks would signify quasiparticles at poles of the single-particle-removal Green's function\cite{Koralek2006}. The strongly decaying spectral intensity approaching \ef also defies conventional behavior in a metal and can be viewed as a pseudogap-like feature persisting to room temperature and distributed over an energy scale of at least several hundreds of meV. Indeed it appears to be associated with the broad-energy, doping-dependent suppression in the density of states at \ef seen in room temperature optical reflectivity experiments on BKBO and BPBO\cite{Karlow1993,Tajima1985}.

Additionally, a second, distinct type of pseudogap-like feature arises at lower temperature. Below \tp, in spectra acquired at 50 K and 17 K, intensity within 100--200 meV of \ef is suppressed relative to the spectra above \tp, while additional weight appears in the tails. Here the energy scale and temperature dependence align with previous angle-\emph{integrated} photoemission measurements\cite{Chainani2001}. The momentum-resolving capability of the present experiments is a key to identifying the origin of this behavior. Though originally dubbed a pseudogap, analysis in the next section will demonstrate that the low-temperature redistribution of spectral intensity is due to the opening of a true bandgap with an associated change in symmetry, albeit one whose origin had been obscured by phase separation.

\section*{Signatures of electronic phase separation}

From a cursory view of Fig.~\ref{fig:ARPES_EDC}, it would be tempting to ascribe the low-temperature spectral buildup in the tails to a simple increase in incoherent states due to, e.g., inelastic scattering from defects in such a highly doped material. But two important observations suggest that the changes are actually associated with the formation of an electronic phase that is distinct from the (bad) metal. First, it is clear from comparing Figs.~\ref{fig:ARPES_EDC}\textbf{c} and \ref{fig:ARPES_EDC}\textbf{d} that the spectral weight redistribution into the tails at low temperature is highly anisotropic in $k$-space. Namely, the relative increase in intensity in the tail along $\Gamma-M$ upon cooling is much less than along $\Gamma-X$. Second, comparing Fig.~\ref{fig:ARPES_EDC}\textbf{c} to the inset of Fig.~\ref{fig:ARPES_EDC}\textbf{d}, which zooms in near \ef along $\Gamma-X$, we can see that the \emph{fraction} of spectral weight lost at \ef at each momentum point upon cooling is nearly the same. 

Further analysis presented in Fig.~\ref{fig:subtraction} demonstrates these observations can be explained by a transition to a superposition of two distinct electronic structures appearing in the ARPES measurements below \tp. In particular, the $k$-isotropic reduction in intensity at \ef results from the fraction of metallic states lost to the breathing-distorted insulator phase. As a result of its symmetry, this phase has dispersive features and spectral weight patterns that differ from those of the metallic band structure, thereby accounting for the $k$-anisotropy of the temperature-dependent spectral tails.

To verify the occurrence of phase separation, we show that the low-temperature spectra can be decomposed in order to isolate the insulating component, $I^{LT}_{ins}(\bm{k},\omega)$, which can then be compared to insulating low- or undoped BKBO. We obtain this component via the subtraction
\begin{equation}\label{eq:subtraction}
I^{LT}_{ins}(\bm{k},\omega)=I^{LT}(\bm{k},\omega)-\alpha I^{HT}(\bm{k},\omega),
\end{equation}
where $I^{LT}(\bm{k},\omega)$ is a spectrum acquired at low temperature,  $I^{HT}(\bm{k},\omega)$ is a high-temperature spectrum obtained well above \tp, and $\alpha$ is a constant equal to the intensity ratio $I^{LT}(k_F,E_F)/I^{HT}(k_F,E_F)$ between high temperature and low temperature scans. By construction, intensity at \ef in the difference spectrum of Eq.~(\ref{eq:subtraction}) will vanish at all points in $k$-space only if $\alpha$, which we extract empirically at an arbitrary \kf, is single-valued on the whole Fermi surface --- a stringent condition that is satisfied in the case of metal-insulator phase separation, as $\alpha$ would then straightforwardly represent the fraction of remaining metallic states at low temperature. 

The outcome of the subtraction is summarized in Fig.~\ref{fig:subtraction}. Starting from the left, the first two columns of Fig.~\ref{fig:subtraction}\textbf{a} present energy-vs.-$k$ data obtained along $\Gamma-M$ from a superconducting sample with doping $x=0.34$, at temperatures well below and above $T_p$ (17 K and 200 K, respectively). The third column is the result of the subtraction operation (Eq.~\ref{eq:subtraction}) between the high and low temperature data using $\alpha=0.6$, while the fourth column (Fig.~\ref{fig:subtraction}\textbf{b}) presents comparison data from a low-doped insulating sample ($x \approx 0.15$, $T = 17$ K). There is a clear similarity between the subtraction spectrum and the insulator in terms of a gap in the spectrum, and the turnaround point of the band at momentum corresponding to the folded Brillouin zone boundary of the insulator (vertical dashed lines). Figure \ref{fig:subtraction}\textbf{c} shows momentum distribution curves (MDCs) at a fixed binding energy of -0.15 eV from each panel in Fig.~\ref{fig:subtraction}\textbf{a} (horizontal dashed lines). These further illustrate how spectral weight is redistributed to higher momentum, such that the gap edge seen in the subtraction spectrum forms around the insulator Brillouin zone boundary.

We have extended this subtraction method over a quadrant of the $\Gamma-M-X$ plane of the Fermi surface obtained by azimuthally scanning the sample. The results are presented in Fig.~\ref{fig:subtraction}\textbf{d} in the form of $k$-space intensity maps at different binding energies. We draw attention to two important aspects of the subtracted data: First, the use of a single scaling factor, $\alpha$, indeed leads to vanishing intensity at \ef over the full Fermi surface. Second, the intensity maps at binding energies of -220 and -300 meV indicate that at low temperature, spectral weight redistributes in $k$-space into the region around the $M$ point (indicated by arrows). As seen from the low-doped $x \approx 0.15$ sample, this is consistent with the folded band structure along $(\pi,\pi,\pi)/a$ and spectral weight distribution of the insulating phase (see further discussion in the Supplementary Information).

\section*{Comparison with a 2D Su-Schrieffer-Heeger model}

\begin{figure}
    \centering
    \includegraphics[width=1.0\textwidth]{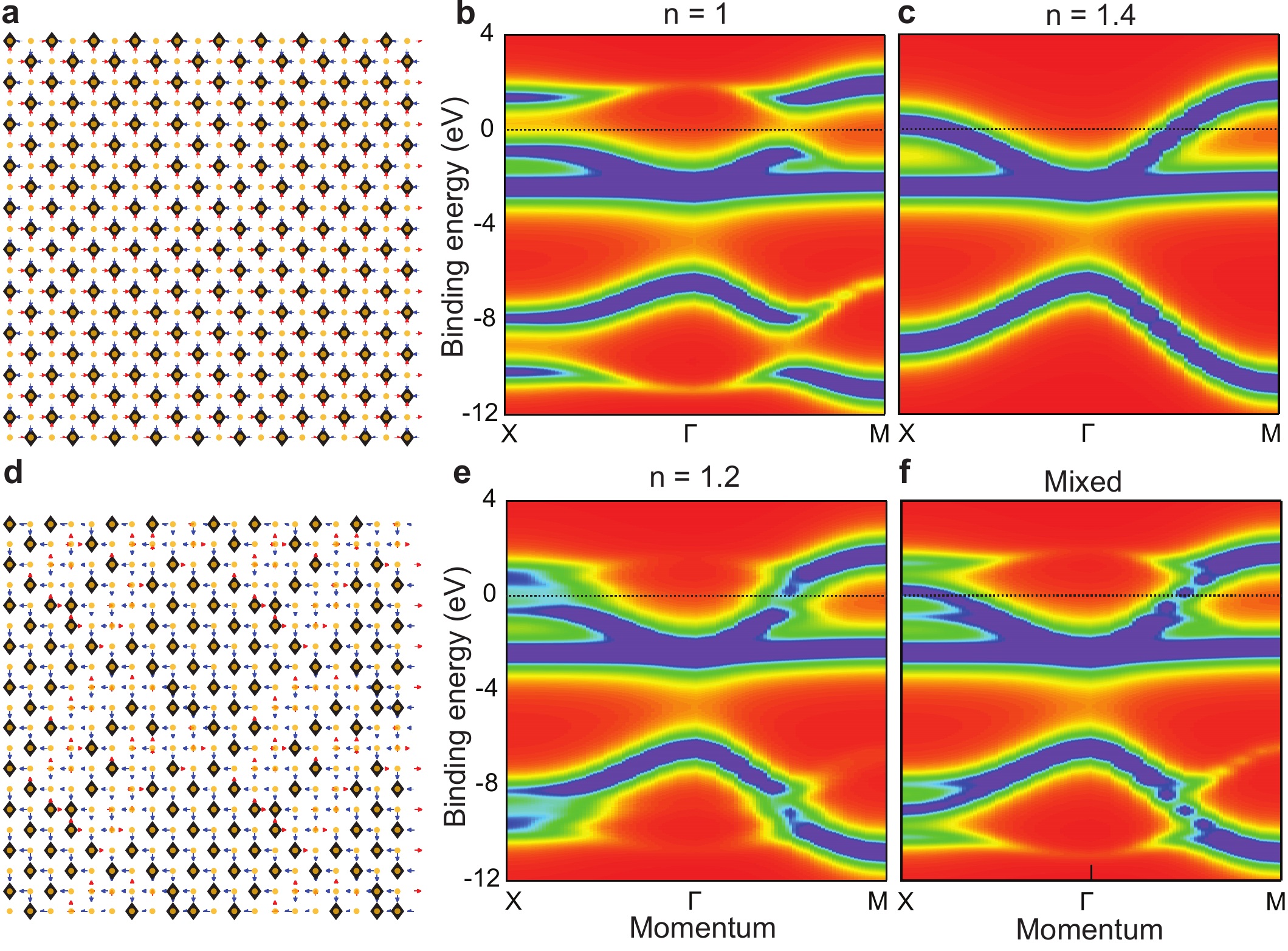}
    \caption{\textbf{Phase separation in a 2D Su-Schieffer-Heeger model.} 
        \textbf{a}, Ground state at half filling ($n=1$). The model yields ordered breathing distortions analogous to 3D BBO. The collapsed BiO$_2$ plaquettes are indicated by black diamonds. Red (blue) arrows represent positive (negative) displacements of the oxygen atoms away from the ideal square lattice.
        \textbf{b}, Insulating band structure of the $n=1$ breathing-distorted phase. 
        \textbf{c}, Calculated band structure for $n=1.4$, which is found to be metallic.
        \textbf{d},\textbf{e} Real-space view and corresponding band structure of the lattice after minimizing the total energy at $n=1.2$.
        \textbf{f}, Mixed spectral function consisting of equal contributions of the $n=1$ insulating and $n=1.4$ metallic spectra from \textbf{b} and \textbf{c}, respectively, for comparison with \textbf{e}.
        } 
    \label{fig:calculation}
\end{figure}

We have compared the findings from ARPES with a 2D cluster calculation based on a three-orbital Su-Schrieffer-Heeger (SSH) model\cite{Su1980} defined on a BiO$_2$ lattice (see Methods). A key aspect of this model is that the electron-lattice coupling modulates the hopping parameters with oxygen displacements. We then minimize the total energy with respect to the static displacement of the oxygen atoms. Although this model neglects factors such as Coulomb interactions and the phonon momenta, it is nevertheless serves to illustrate how short-range electron-phonon interactions can drive the formation of insulating bipolaron order at low doping and intrinsic phase separation at intermediate doping. As shown in Fig.~\ref{fig:calculation}\textbf{a}, at a density $n$ of one hole per Bi, the model yields a breathing-distorted ground state analogous to three-dimensional \bbo, which opens a gap in the conduction band (Fig.~\ref{fig:calculation}\textbf{b}). By contrast, $n=1.4$ is metallic (Fig.~\ref{fig:calculation}\textbf{c}). At intermediate doping --- in this case $n=1.2$ --- nanoscale regions of ordered breathing distortions cluster together, separated by areas of irregular Bi-O distortions (Fig.~\ref{fig:calculation}\textbf{d}). Although the model used here is a toy model for the real system, the resulting spectrum shows evidence of superimposed metallic and insulating band structures (Figs.~\ref{fig:calculation}\textbf{e}, \textbf{f}), supporting the conclusions from of our analysis of the ARPES data. We hope that these observations will motivate further theoretical studies of related models. In particular, the current model as constructed may undergo spinodal decomposition (as strongly suggested by the numerical results), in which case the true ground state should be expected to have very large regions of metal or insulator. A more realistic model would take long elastic couplings and/or Coulomb constraints into account to obtain finite size bubbles of phase separated regions consistent with experiments.

\section*{Temperature dependence of Raman and ARPES data}

Since insulating behavior in the BKBO phase diagram is fundamentally connected with ordered BiO$_6$ breathing distortions, we have compared the temperature dependence of the electronic and atomic structure to gain insights into the nature of the transition to electronic phase separation occurring across \tp. Figure \ref{fig:Raman}\textbf{a} plots ARPES difference spectra along $\Gamma-M$ obtained by Eq.~(\ref{eq:subtraction}). Here the high-temperature reference spectrum $I^{HT}(\bm{k},\omega)$ was measured at 255 K, while the various low-temperature spectra $I^{LT}(\bm{k},\omega)$ were measured at the temperatures indicated in the figure.  A clear change occurs below $\sim160$ K, at which point band weight can be seen dispersing up to about -200 meV binding energy, reaching the folded Brillouin zone boundary of the insulating phase --- the signature of insulating band structure and phase separation established in Fig.~\ref{fig:subtraction}.

It is interesting to compare the relatively well-defined temperature range of this electronic phase separation with the temperature dependence of the local atomic structure. Although diffraction experiments report that BKBO in this doping range has --- globally, at least --- simple cubic\cite{Pei1990} (or tetragonal\cite{Braden2000}) structure, Raman scattering, which is more sensitive to local structure, has found evidence of persistent BiO$_6$ breathing distortions\cite{Tajima1992,Menushenkov2003}. The discrepancy between diffraction and Raman measurements signals that the breathing distortions are highly disordered and/or fluctuating. Raman spectra presented in Fig.~\ref{fig:Raman}\textbf{b} confirm the presence of breathing distortions in our own thin film samples, as indicated by the peak located at an energy shift of 564 cm$^{-1}$. The peak, which shows an asymmetric Fano-like broadening that is likely due to coupling with the electrons, is visible at all temperatures. It shows little qualitative change as a function of temperature, save for a slight sharpening below \tp of the subtle contours of the broadened structure (blue curves) with respect to temperatures in the phase separation transition region (green curves) and above it (red curves).

The stability of the Raman features as a function of temperature suggests that the mere existence of breathing distortions is not the sole factor at play in the metal-insulator phase separation at \tp; presumably some lengthscale of both structural and electronic coherence needs to be established. Indeed, we find evidence from ARPES suggesting that electronic scattering is a key determinant in restricting the formation of the the insulating regions.  The left axis of Fig.~\ref{fig:Raman}\textbf{c} shows the Lorentzian half-width at half-maximum, $\gamma$, of the metallic band at \ef as a function of temperature. The values are obtained from fits to MDCs along the $\Gamma-X$ direction.  On the right axis we plot the insulator-associated signal intensity obtained from the difference spectra of Fig.~\ref{fig:Raman}\textbf{a}. This ``insulator weight'' comes from integrating counts in a region focused on the insulating band, indicated by the white lines in the top left image of Fig.~\ref{fig:Raman}\textbf{a}, and is adopted as a metric to characterize the extent of the electronic transition.

In ARPES, $\gamma$ is equal to the inverse of the mean free path of the electrons, $\xi^{-1}$. The presented measurements of the low-energy electrons near the $X$ point are of particular interest, as these states are the closest on the Fermi surface to a van Hove singularity\cite{Sahrakorpi2000}. At the highest measured temperature of 255 K, the MDC linewidth corresponds to an extremely high scattering rate: On average, electrons at \ef scatter after traveling only about $1.7a$, where $a$ is the cubic lattice constant (4.29 \AA). This is close to the Mott-Ioffe-Regel limit, below which localization would be expected. Tracking the insulator weight while decreasing temperature, we see an onset of the transition at about 160 K and completion near 110 K. At the same time, $\gamma$ decreases monotonically. We observe that the transition centered at \tp occurs in a temperature range where $\xi$ begins to exceed $2a$, the characteristic lengthscale regime of the ordered breathing distortions in insulating BKBO, where the cubic unit cell doubles along each crystal axis. 

Given that extrinsic effects can broaden the measured linewidths, these values of $\xi$ deserve to be thoroughly scrutinized. Ultimately, we find compelling evidence that the linewidths and corresponding values of $\xi$ are close to intrinsic (see Supplementary Information). For example, we have repeated the experiments at different photon energies to test for the influence of  perpendicular momentum broadening and surface effects in the photoemission process. As shown in Fig.~\ref{fig:Raman}\textbf{c}, we obtain the same $\xi(T)$ results using photon energies of 70 eV and 120 eV. 

Moreover, the notion that the phase separation is regulated by a temperature-controlled crossover in $\xi$ qualitatively accounts for the otherwise difficult-to-explain doping dependence of \tp seen in transport measurements. Having now found a correlation between the resistivity elbow and the phase separation transition at \tp, it appears that \tp \emph{increases} with higher doping until its disappearance at $x \approx 0.4$ (ref.~\citen{Nagata1999}). The same qualitative shift in \tp can be seen in our own thin film samples (Fig.~\ref{fig:intro}\textbf{b}). At first glance, this trend is deeply counterintuitive; it would be natural to assume that \tp should decrease as doping is increased in the range of about $x=0.35$ to 0.4, going away from the insulator phase below $x \approx 0.3$. The observed behavior can be explained, though, in the scattering-regulated scenario. Scattering appears to decrease with increasing $x$ as the samples move toward a more Fermi-liquid-like transport regime, which is not surprising, as both charge correlations and electron-phonon coupling would be reduced at higher doping. Higher $x$ therefore implies longer electronic mean free paths, which in our picture leads to higher \tp.

\section*{Discussion}

Depending on one's definition, our data can be interpreted in terms of two types of pseudogaps in BKBO. The first exists up to at least room temperature and takes the form of a broad suppression of spectral weight and absence of clear quasiparticles approaching \ef. The second is the energetically sharper near-\ef intensity loss below \tp. We have shown that, in reality, this second pseudogap could alternatively be regarded as a true bandgap, albeit one whose origins and underlying symmetry had been obscured by phase separation.

Our results indicate that these two pseudogap-like features and the transition between them stem from (bi)polaron interactions. First and most obviously, the band structure of the insulating fraction below \tp reflects folding present in low-/undoped BKBO attributed to ordered bipolarons. Secondly, this insulating band structure is formed out of a spectral weight transfer from near-\ef states into deep-energy tails that already existed at temperatures above \tp. Combined with Raman results showing the presence of disordered and/or fluctuating breathing distortions at high temperatures, this spectral weight transfer and the anomalous quasiparticle-less lineshapes that precede it give the impression that incoherent states of the bad metal phase (or, rather, subregions of it) are itinerant precursors to the localized, ordered bipolarons that precipitate below \tp. In this view, energetics favorable to bipolarons would already be established above \tp, but scattering  precludes their phase condensation into an ordered phase --- a scenario that fits well with our observation that \tp is correlated with a crossover in the electronic mean free path. 

The dispersive band trailed by a large incoherent tail seen here in BKBO bears similarities to ARPES measurements of other materials thought to show hallmarks of polarons, such as low-doped TiO$_2$\cite{Moser2013}, SrTiO$_3$\cite{Wang2016}, and manganites\cite{Sun2006,Massee2011}. BKBO shows a particularly striking resemblance to $R$NiO$_3$ rare earth nickelates. Those compounds also undergo metal-insulator phase transitions, and there are claims that they have a bond-disproportionated ground state analogous to the BKBO parent compound\cite{Park2012,Johnston2014,Bisogni2016}. ARPES data from NdNiO$_3$ show spectral features (weak, though observable, quasiparticle-like peaks and strong deep-energy tails) that are qualitatively similar to BKBO\cite{Schwier2012,Dhaka2015}. Moreover, as in BKBO, decreasing temperature across the metal-insulator transition in NdNiO$_3$ leads to a transfer of spectral weight from near the Fermi level into the deep-energy tails\cite{Schwier2012,Dhaka2015}. In contrast to BKBO, however, NdNiO$_3$ is undoped, and photoemission intensity at \ef vanishes below the transition temperature; the sample becomes purely insulating, rather than phase separated. Similar to the results from Raman spectroscopy on BKBO, measurements of the pair distribution function (PDF) in nickelates by neutron scattering have revealed local distortions related to the insulating phase that are ``hidden'' within the global symmetry and persist above the metal-insulator transition temperature\cite{Li2016,Shamblin2018}. Based on both PDF and dielectric spectroscopy measurements, Shamblin \emph{et al}.~have proposed a scenario similar to the one here, in which the metal-insulator transition in nickelates represents the freezing of a polaronic liquid\cite{Shamblin2018}.

In all the examples just mentioned, however, EDC peaks of the metallic bands signal that at least some fraction of coherent spectral weight survives the polaron interactions, whereas electrons in BKBO are scattered to an even greater extent, such that we see no peaks. The observation of a dispersive band persisting in the absence of coherent spectral weight runs counter to Fermi liquid theory, which predicts that effective mass should become infinite as the quasiparticle residue, $Z_{\bm{k}}$, goes to zero. The behavior may be in line, though, with the SSH model, in which electron-phonon coupling modulates the hopping parameters, rather than the local potential via the charge density. Recent theoretical work on such a model found light bipolarons can exist, even in the strong coupling regime\cite{Sous2018}.

Metal-insulator phase separation is potentially an important factor influencing superconductivity in BKBO. An obvious point is that phase separation sidesteps a global metal-insulator transition, as occurs in the nickelates, allowing for the survival of a metallic Fermi surface amid strong interactions and setting the stage for superconductivity.  At a deeper level, recent studies have highlighted how disorder\cite{Harris2018} and the lengthscale of structural phase separation\cite{Giraldo-Gallo2015} in the bismuth oxides appear to be relevant to \tc. Additionally, the electronic phase mixture seen above \tc implies that a direct superconductor-insulator transition (SIT) indeed occurs along the $x$ axis of the phase diagram at $T=0$, as has long been suspected. This positions BKBO as a platform for the study of SITs, and it may have important consequences for the nature of its superconducting pair state\cite{Trivedi2012b}.

\begin{methods}

\subsection{Sample preparation}
Samples were prepared at the Paul Scherrer Institut. Thin films were grown on SrTiO$_3$(001) substrates by pulsed laser deposition. Ablation targets were prepared by a two-step solid state synthesis. Stoichiometric mixtures of KO$_2$, Bi$_2$O$_3$, and pre-synthesized BaBiO$_3$\cite{Plumb2016} were well mixed into a paste inside a helium glove box, then annealed at 725 $^\circ$C in dynamic vacuum for 1 h, and finally furnace cooled. Subsequently, the material was heated up to 425 $^\circ$C, kept at this temperature for 1 h, and slowly cooled down at a rate in the range of 5-15 $^\circ$C/h. The material was then further annealed for 1 h at 400 $^\circ$C, 375 $^\circ$C, 350 $^\circ$C, and 325 $^\circ$C with slow cooling, all in flowing oxygen atmosphere. For each batch of the material, the whole procedure was repeated three times. The resulting powder was pressed into pellets with a load of 15 tons. Pellets were sintered in oxygen at temperatures of 425--325 $^\circ$C, as previously applied for the powder synthesis. 

The film growth was performed using an Nd:YAG laser with a repetition rate of 2 Hz and pulse fluence of 1.6 J/cm$^{2}$. Films were grown on SrTiO$_3$(001) substrates in 0.5 mbar of oxygen with substrates held at a temperature of 480 $^\circ$C. Two-dimensional epitaxial growth was verified by reflection high-energy electron diffraction (RHEED). After deposition, the films were annealed at 370 $^\circ$C in 1 bar of pure oxygen for 0.5 hours. Films were roughly 11 nm thick. For the ARPES measurements, films were transfered directly from the deposition chamber to the experimental station via an ultrahigh vacuum connection with pressure better than $5 \times 10^{-9}$ mbar. In addition to being studied by ARPES and Raman scattering, samples were characterized by x-ray diffraction, x-ray fluorescence, x-ray photoelectron spectrsocopy, and resistivity measurements. See the Supplementary Information for more details.

\subsection{Angle-resolved photoemission spectroscopy} 
ARPES experiments were performed at the Surface/Interface Spectroscopy (SIS) beamline X09LA of the Swiss Light Source. The endstation is equipped with a hemispherical electron analyzer and 6-axis cryogenic manipulator. The presented data were collected using $p$-polarized light. The total energy resolution was 10 meV. The hemispherical analyzer has an angular resolution of 0.1$^\circ$. Momentum-space intensity maps were acquired by scanning the azimuthal rotation axis at fixed tilt and polar angles. The pressure during the measurements was better than $5 \times 10^{-11}$ mbar. 

\subsection{Raman scattering}
Raman scattering experiments were performed at the University of Geneva using a homemade micro-Raman setup based on a half-meter spectrometer coupled to a nitrogen-cooled Princeton Instruments CCD detector. A gas laser at a wavelength of 514.5 nm was used for excitation. Narrow edge filters allowed measuring Stokes lines down to 50 cm$^{-1}$. A long working distance $\times63$ objective (N.A. 0.7) was used. In order to avoid overheating, we used a very low laser power (less than 100 $\mu$W) for a spot size of about 2 $\mu$m in diameter. For the temperature dependent measurements, the samples  were mounted on a Cryovac helium flow cryostat. Full spectra were acquired at identical temperatures for both the BKBO films and a bare SrTiO$_3$ substrate. The Raman response of the BKBO thin film presented in Fig.~\ref{fig:Raman}\textbf{b} was obtained after subtraction of the substrate contribution.

\subsection{Theoretical model} 
We define a three-orbital Su-Schrieffer-Heeger model on a 2D Lieb lattice. The orbital basis consists of a Bi $6s$ atom and two O $2p$ orbitals situated halfway between each of the Bi atoms. The positions of the heavier Bi atoms are fixed, and O atoms are restricted to move along the bond direction. The Hamiltonian is

\begin{equation*} \label{eq:Hamiltonian}
\begin{split}
H = & -t_{sp} \sum_{\rm{r},\sigma} (1-\alpha x_{\mathbf{r}}) (s_{\mathbf{r},\sigma}^\dagger p_{\mathbf{r} ,x,\sigma}+h.c.)
      -t_{sp} \sum_{\mathbf{r},\sigma} (1-\alpha y_{\mathbf{r}}) (s_{\mathbf{r},\sigma}^\dagger p_{\mathbf{r},y,\sigma}+h.c.) \\
    & +t_{sp} \sum_{\mathbf{r},\sigma} (1+\alpha x_{\mathbf{r}}) (s_{\mathbf{r+a},\sigma}^\dagger p_{\mathbf{r},x,\sigma}+h.c.)  
      +t_{sp} \sum_{\mathbf{r},\sigma} (1+\alpha y_{\mathbf{r}}) (s_{\mathbf{r+b},\sigma}^\dagger p_{\mathbf{r},y,\sigma}+h.c.) \\
    & +t_{pp} \sum_{\mathbf{r},\sigma} (p_{\mathbf{r},x,\sigma}^\dagger p_{\mathbf{r},y,\sigma} - p_{\mathbf{r},y,\sigma}^\dagger p_{\mathbf{r-a},x,\sigma}
                                       +p_{\mathbf{r-a},x,\sigma}^\dagger p_{\mathbf{r-b},y,\sigma} - p_{\mathbf{r-b},y,\sigma}^\dagger p_{\mathbf{r},x,\sigma}) \\
    & +\sum_{\mathbf{r},\sigma} (\epsilon_{s} \hat{n}_{\mathbf{r},\sigma}^{s} + \epsilon_{p} \hat{n}_{\mathbf{r},\sigma}^{p_{x}} 
      + \epsilon_{p} \hat{n}_{\mathbf{r},\sigma}^{p_{y}}) 
      + \sum_{\mathbf{r}} (K x_\mathbf{r}^2+K y_\mathbf{r}^2).
\end{split}
\end{equation*}

Here, the operators $s_{\mathbf{r},\sigma}^\dagger$ ($s_{\mathbf{r},\sigma}$) and $p_{\mathbf{r},\sigma}^\dagger$ ($p_{\mathbf{r},\sigma}$) are the creation (annihilation) for the spin $\sigma$ holes on the Bi 6s and O 2p orbitals respectively. The unit cells are indexed by $\mathbf{r} = n_x \mathbf{a} + n_y \mathbf{b}$, where $(n_x, n_y) \in \mathbb{Z}$, $\mathbf{a} = (a, 0)$ and $\mathbf{b} = (0, a)$ are the primitive lattice vectors along the $x$ and $y$ directions, respectively, and $a$ is the Bi-Bi bond length of the undistorted lattice. The operators ${n}^{s}_{\mathbf{r},\sigma} = {s}^{\dagger}_{\mathbf{r},\sigma}s_{\mathbf{r},\sigma}$ and 
${n}^{p_{\delta}}_{\mathbf{r},\sigma} = {p}^{\dagger}_{\delta,\mathbf{r},\sigma}p_{\delta,\mathbf{r},\sigma}$ are the number operators for $s$ and $p_{\delta}$
$(\delta = x, y)$ orbitals, respectively. The displacement of oxygen atoms is described by $x_{\mathbf{r}}$ and $y_{\mathbf{r}}$, and the electron-phonon
coupling modulates the hopping integral $t_{sp}$ by $\alpha x_{\mathbf{r}}$ and $\alpha y_{\mathbf{r}}$. $K$ is the coefficient of elasticity between each Bi
and O atoms, and each O atom is linked by two springs to the neighboring Bi atoms. We have neglected the kinetic
energy term for oxygen atoms.

The solution of the Hamiltonian is found by searching for $\frac{\partial E}{\partial x_{\mathbf{r}}}=0$ and $\frac{\partial E}{\partial y_{\mathbf{r}}}=0$, where $E$ is the energy of the system given by 
\begin{equation*}
    E = \sum_{m,\sigma}f(E_{m,\sigma})E_{m,\sigma}=\sum_{m,\sigma}f(E_{m,\sigma})\expval{H}{\Psi_{m,\sigma}},
\end{equation*}
where $E_{m,\sigma}$ is the eigenenergy for spin $\sigma$ corresponding to the $\Psi_{m,\sigma}$ eigenstate.  $f(E_{m,\sigma})$ is the Fermi-Dirac distribution with respect to $E_{m,\sigma}$. 

We adopt hopping parameters motivated by density functional theory calculations\cite{Foyevtsova2015} with $t_{sp} = 2.08$ eV, $t_{pp} = 0.056$ eV, $\epsilon_s = 6.42$ eV, and
$\epsilon_p = 2.42$ eV. The \emph{e-ph} coupling constant is $\alpha = 4a^{-1}$ and $K=0.104$ $\mathrm{eV}/a^2$, which leads to a distortion of $0.075a$ at half-filling. Calculations were performed on a large cluster with $N = 20 \times 20$ cell size. The spectral weight function was calculated via
\begin{equation*}\label{eq:AKW}
A(\mathbf{k},\omega) = \frac{1}{2\pi} \sum_{m,\gamma,\sigma} \frac{\mid \langle \mathbf{k}, \gamma \mid \Psi_{m,\sigma} \rangle \mid ^2}{\omega -E_{m,\sigma}+{\mathrm i} \delta},
\end{equation*}
where $\delta=0.3$ eV, $\gamma$ is the orbital index and $\mid \mathbf{k},\gamma\rangle = \frac{1}{N} \sum_{\mathbf{r}} e^{{\mathrm i}\mathbf{k\cdot r}}\mid\gamma_{\mathbf{r}}\rangle$.

\end{methods}

\bibliography{citations}


\begin{addendum}
 \item The authors are grateful for thought-provoking discussions with T.M.~Rice, and J.~Chang. A.~Pfister and L.~Nue lent technical assistance at SIS beamline. C.W.~Schneider assisted with instrumentation in the laboratory of the Thin Films and Interfaces group at PSI. M.N.~is supported by the Swiss National Science Foundation under project 200021\_159678. This project has received funding from the European Union's Horizon 2020 research and innovation programme under the Marie Sk\l{}odowska-Curie grant agreement No 701647. D.J.G. received financial support from SCIEX NMS\textsuperscript{CH} (Project No. 13.236) granted by the Rectors Conference of the Swiss Universities.
A portion of the work was conducted at the Center for Nanophase Materials Sciences, which is a DOE Office of Science User Facility. T.B. and S.J. acknowledge support from the Scientific Discovery through Advanced Computing (SciDAC) program funded by the U.S. Department of Energy, Office of Science, Advanced Scientific Computing Research and Basic Energy Sciences, Division of Materials Sciences and Engineering.
 
 \item[Author contributions] M.N.~grew the BKBO films and performed ARPES with help from M.Y., J.J., J.M., M.R.~and N.C.P.. Z.R.~helped with studies of optimal film growth conditions. D.G.~produced the ablation targets for the film growth under the supervision of M.Medarde and E.P.. J.T.~and A.S.~performed temperature-dependent Raman measurements under the supervision of D.v.d.M.. M.N.~characterized the films by various techniques (resistivity, x-ray diffraction, and x-ray fluorescence) with help from T.S., D.G., and E.P.. S.L., T.B., and S.J. performed the cluster calculations. M.N.~analyzed the data with valuable feedback from N.C.P., M.M\"{u}ller, S.J., T.B., M.S., M.R., and J.M.. M.N.~and N.C.P.~wrote the manuscript with input from all the coauthors. N.C.P.~supervised the project and conceived it together with M.S.~and M.R. All authors discussed the results.
 \item[Competing Interests] The authors declare that they have no
competing financial interests.
 \item[Correspondence] Correspondence and requests for materials
should be addressed to M.N.~(email: muntaser.naamneh@psi.ch) or N.C.P.~(email: nicholas.plumb@psi.ch).
\end{addendum}

\newpage

\newpage

\begin{figure}
\centering
\includegraphics[width=0.95\textwidth]{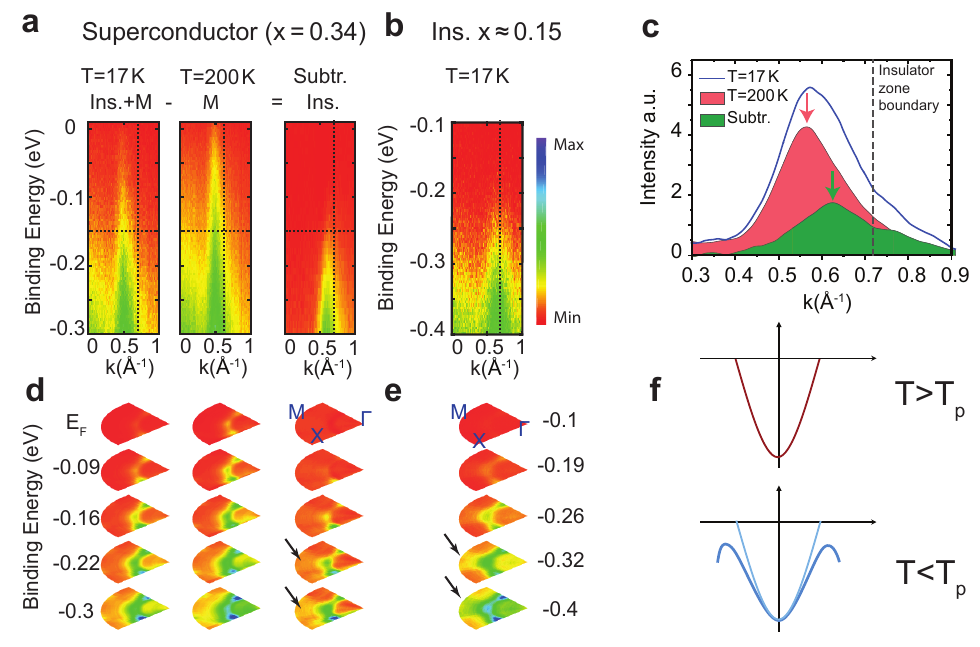}
\end{figure}

\newpage

\begin{figure}
    \centering
    
    \caption{\textbf{Decomposition of ARPES spectra revealing temperature-driven metal-insulator phase separation in BKBO.} 
        \textbf{a}, Energy-vs.-$k$ spectra along the $\Gamma-M$ direction of a superconducting sample with $x=0.34$. The left and middle panels were measured at temperatures of 17 K and 200 K, respectively. The right panel is obtained by the subtraction method of Eq.~\ref{eq:subtraction}, taking the 17 K and 200 K spectra as $I^{LT}(\bm{k},\omega)$ and $I^{HT}(\bm{k},\omega)$, respectively.
        \textbf{b}, Analogous data obtained at 17 K from low-doped, insulating BKBO ($x \approx 0.15$). The binding energy range is shifted to account for an offset in the chemical potential, which might result from the varying amount of hole states trapped inside the gap by polaronic effects\cite{Franchini2009}. The vertical dashed lines in \textbf{a} and \textbf{b} show the location of the Brillouin zone boundary in the \emph{insulating} phase. 
        \textbf{c}, MDCs at a binding energy of -0.15 eV from the spectra in \textbf{a}, as indicated by the horizontal dashed lines in those panels.
        \textbf{d}, \textbf{e}, Similar to \textbf{a} and \textbf{b}, but showing momentum-space intensity maps at different binding energies. 
    The arrows highlight a signature of the insulating phase in \textbf{d} --- weight around the $M$ point --- that is also visible in the subtracted images of \textbf{c}.
        \textbf{f}, Sketch of the evolution of the spectrum with temperature. A single metallic band is seen at $T>T_p$, while a superposition of a weaker metallic band and insulating band exists for $T<T_p$.
    }
    \label{fig:subtraction}
\end{figure}

\newpage

\newpage

\begin{figure}
    \centering
    \includegraphics[width=0.95\textwidth]{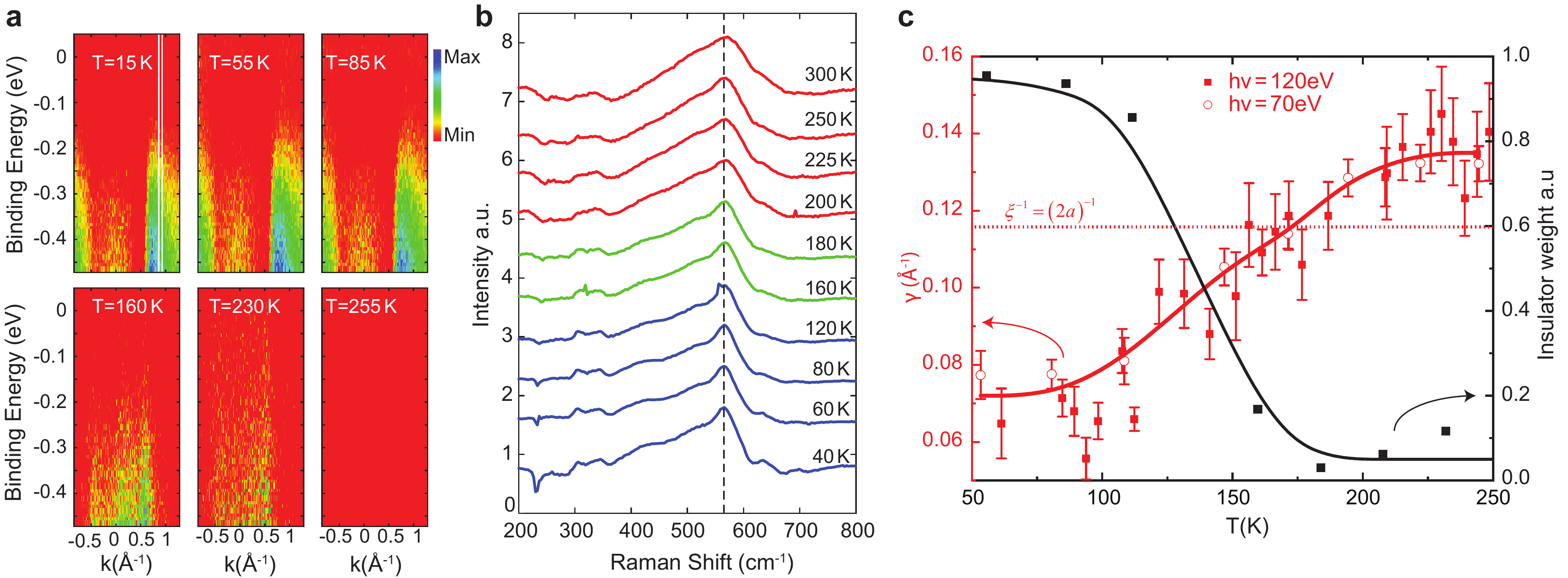} 
    \caption{\textbf{Temperature dependence of the electronic phase separation.} 
        \textbf{a}, ARPES difference spectra along the $\Gamma-M$ direction for various labeled temperatures. The spectra are calculated by the method applied in Fig.~\ref{fig:subtraction} and described in the supporting text. All difference spectra are taken with respect to 255 K. The high intensity features seen when $T<T_p$ (top row) are signatures of the insulating band structure.
        \textbf{b}, Temperature-dependent Raman spectra. The breathing mode feature peaked at 564 cm$^{-1}$ (dashed line) is present at all temperatures and shows a strong asymmetric Fano-like broadening.
        \textbf{c}, Left axis: Fitted Lorentzian half-width at half-maximum, $\gamma$, of  momentum distribution curves (MDCs) of the metallic band at \ef. The data were acquired from different samples using two different incoming photon energies, 70 eV and 120 eV. For 70 eV, the MDC is nearly along the $\Gamma-X$ line. The 120 eV data was collected in the same geometry in order to test the robustness of the results against $k_z$ broadening and surface effects. The horizontal dashed line indicates where $\gamma$ corresponds to an electron mean free path, $\xi$, equal to $2a$ --- the characteristic lengthscale of the ordered insulating phase. Right axis: Intensity of the insulating band. The values are obtained by summing the difference spectra in the window region indicated by vertical white lines in the top left image of \textbf{a}. The curves are hand-drawn guides to the eye.}
    \label{fig:Raman}
\end{figure}

\end{document}